\documentclass[journal,twoside,web,UTF8]{ieeecolor}
\usepackage{generic}
\usepackage{cite}
\usepackage{amsmath,amssymb,amsfonts}
\usepackage{algorithmic}
\usepackage{algorithm}
\usepackage{graphicx}
\usepackage{textcomp}
\usepackage{booktabs}
\usepackage{amssymb}
\usepackage{caption}
\usepackage{subfigure}
\def\BibTeX{{\rm B\kern-.05em{\sc i\kern-.025em b}\kern-.08em
    T\kern-.1667em\lower.7ex\hbox{E}\kern-.125emX}}
\begin{document}
\title{Intelligent Sensing Scheduling for Mobile Target Tracking Wireless Sensor Networks}
\author{Longyu Zhou, Supeng Leng, \IEEEmembership{Member, IEEE}, Qiang Liu, Haoye Chai, and Jihua Zhou, \IEEEmembership{Member, IEEE} 
}

\maketitle

\begin{abstract}
Edge computing has emerged as a prospective paradigm to meet ever-increasing computation demands in Mobile Target Tracking Wireless Sensor Networks (MTT-WSN). This paradigm can offload time-sensitive tasks to sink nodes to improve computing efficiency. Nevertheless, it is difficult to execute dynamic and critical tasks in the MTT-WSN network. Besides, the network cannot ensure consecutive tracking due to the limited energy. To address the problems, this paper proposes a new hierarchical target tracking structure based on Edge Intelligence (EI) technology. The structure integrates the computing resource of both mobile nodes and edge servers to provide efficient computation capability for real-time target tracking. Based on the proposed structure, we formulate an energy optimization model with the constrains of system execution latency and trajectory prediction accuracy. Moreover, we propose a long-term dynamic resource allocation algorithm to obtain the optimal resource allocation solution for the accurate and consecutive tracking. Simulation results demonstrate that our algorithm outperforms the deep Q-learning over 14.5\% in terms of system energy consumption. It can also obtain a significant enhancement in tracking accuracy compared with the non-cooperative scheme.     
\end{abstract}


\section{Introduction}
\label{sec:introduction}
\IEEEPARstart{A}{long} with the penetration of Artificial Intelligence (AI), smart Internet of Things (IoT) system is evolving as an emerging paradigm to facilitate the development of smart city, intelligent agriculture, and intelligent healthcare \cite{Zheng,Wang,JChen, LZhang}. As a kind of attractive IoT applications, Mobile Target Tracking Wireless Sensor Networks (MTT-WSN) has made contributions in many fields, such as illegal vehicle tracking, frontier security, pasturing protection, plant district security, and space exploring. 

The MTT-WSN comprised by many static or mobile nodes can track mobile targets based on the IoT technology. The essential sensor nodes can be activated around the monitoring target during the target tracking, whereas other nodes can be turned into a dormant state to reduce system energy consumption. Take civil aviation flight as an instance, sensor nodes can be activated to drive adverse obstructs including flying birds for ensuring the safety of flight take-off and landing. Based on the sensing data, the MTT-WSN can estimate the status of targets to make real-time tracking decisions. 

The inherent characteristics of MTT-WSN affect the tracking performance, such as the limited sensing ranges and the scarce energy \cite{MWang}. Nevertheless, many static sensors are typically deployed randomly with stationary sensing capability. The deployment manner may cause detection failure due to the blind detection zones, or redundant sensing data when coverage zones overlap. The redundant data can also make bandwidth resource dissipation during data transmission. Moreover, the prediction accuracy cannot be ensured because of the detection failure. In this case, the prediction error can be accumulated so that sensor nodes cannot be activated and scheduled correctly. In addition, the limited energy of sensor nodes cannot assure consecutive target tracking when the mobile targets invade with high moving speed \cite{GQiao,Zhou}. 
 
In fact, the tracking performance is coupled with resource scheduling decisions which can be implemented based on cloud computing or edge computing. Although cloud computing can provide sufficient computing resource, it causes high transmission latency because of long round trip of information delivery \cite{Drossu}. The edge computing appears to reduce the latency while resulting in tremendous computing pressure on edge servers. Besides, it may not be able to integrate available resource to perform real-time feedback for time-critical missions. Emerging AI technology can alleviate the above-mentioned disadvantages on resource scheduling \cite{Liang, JJJ, CCC, B, XiongKai}. However, it is also difficult to design AI based schemes for real-time and consecutive tracking while keeping low communication and computing overheads.

In this paper, we propose a new hierarchical target tracking structure for consecutive target tracking. Based on this structure, an intelligent cloudlet pattern is designed that is composed of edge servers and MNs, in which the MNs can eliminate the redundant sensing data to reduce the latency of data transmission. Besides, the pattern can realize accurate tracking by integrating the computing resource of edge servers and MNs. Moreover, MNs not only can implement target tracking by the trajectory prediction, but also can activate SNs to observe the invading targets collaboratively. The main contributions are summarized as follows. 

\begin{itemize}
\item[\textbf{•}] We first propose a new hierarchical target tracking structure, in which mobile nodes realize the full-scale area coverage with flexible mobility. Besides, the mobile nodes can coordinate the sensing resource of static nodes to observe targets collaboratively. To obtain efficient sensing data, a multi-resource information fusion scheme is proposed to reduce redundancy of data for the maximal sensing resource utilization. 
       
\item[•] Based on the edge intelligence technology, an intelligent cloudlet pattern is proposed to ensure accurate tracking. The pattern integrates the computing resource of both mobile nodes and edge servers to improve the accuracy of trajectory prediction. Moreover, the edge servers can estimate the status of targets to improve computation efficiency for the accurate tracking performance. 
		
\item[\textbf{•}] To realize consecutive tracking under the node scheduling, we present a Long-Term Dynamic Resource Allocation (LTDRA) algorithm. The algorithm can enhance the self-learning nature of traditional reinforcement learning algorithm to explore the optimal decision with the minimal energy consumption and quick algorithmic convergence. In this case, the optimal tracking scheduling can be obtained for the implementation of consecutive tracking. 
	\end{itemize}

The rest of this paper is organized as follows. The related work is given in Section \uppercase\expandafter{\romannumeral2}. Section \uppercase\expandafter{\romannumeral3} gives the system structure and the problem formulation. The edge intelligence framework is proposed in Section \uppercase\expandafter{\romannumeral4}. The evaluation results are provided in Section \uppercase\expandafter{\romannumeral5}. Finally, Section \uppercase\expandafter{\romannumeral6} concludes this paper.

\section{Related Work}
In recent years, edge computing has attracted significant attention in MTT systems \cite{Chiang}. For instance, Kuo. \textit{et al.} proposed an adaptive mechanism of trap coverage with a robust area coverage model in target tracking and services detection to reduce the target-missing time. For merging the cooperation among sensors seamlessly, much work is focused on the collaborative management of computing and moving\cite{Kuo}. A collaborative sensor movement algorithm was proposed on a basis of target learning to minimize the energy consumption\cite{Didd}. Wan \textit{et al.} provided a joint range-Doppler-angle estimation solution for intelligent tracking to improve the efficiency of multi-target tracking\cite{Wan}.

With the development of AI, many researches have studied AI-enabled edge computing \cite{Dai, Bader, Sharma}. For instance, the authors in the literature \cite{Bader} proposed a novel concept to vest the front-end intelligently to realize low-latency in large scale application-oriented IoT scenario. In order to provide real-time information and feedback to the end-users, Sharma introduced a distributed framework for the coordinated process between Mobile Edge Computing (MEC) and cloud computing \cite{Sharma}.

Collaborative computing provides a new opportunity for resource allocation in target tracking applications. Campbell M E \textit{et al.} proposed a cooperative tracking approach for uninhabited aerial vehicles (UAVs) with camera-based sensors, utilizing a square root sigma point information filter, which brought important properties for numerical accuracy, tracking accuracy, and fusion ability \cite{Campbel}. Kuang \textit{et al.} presented a collaborative computational framework that was capable of dealing with many real-world visual tracking problems. A novel spatio-temporal weighting scheme was introduced to maximize the separation between target and background, improving classification accuracy \cite{YZhang}. In mobile bionanosensor network, Okaie \textit{et al.} proposed a cooperative scheme that the bacterium-based autonomous biosensors released repellents to quickly spread over the environment for searching target and release attractants to recruit other biosensors in the environment toward the location around the target for detecting target \cite{Lai}.

\begin{table*}[htbp]
\caption{\label{tab:test}Description and Definition of Symbols} 
	\centering
	\begin{tabular}{|c|l|c|l|c|l|} 
  \hline 
  Notation &  Description & Notation & Description \\ 
  \hline
$L$ & The size of task execution &$\tau_{ar}$ & The task executing deadline\\
\hline
$\mathcal{A^L}$ & task offloading decision space of $a_l$ &$\mathcal{P}$  & The computing set with multi-element integration $\mathcal{F}$\\  
\hline
$r_{i,j}^{{S}\xrightarrow{}{}{M}}$ & The transmission rate from target to the \textit{j}-th MN& $r_{i,s}^{{S}\xrightarrow{}{}{s}}$ & The transmission rate from the \textit{i}-th SN to the sink node\\
\hline
$\tau_c$, $\tau_\beta$ & The offloading latency and computing latency & $f^e$ & Intelligent computing frequency\\
\hline
$x_t$ & Location information of target node at time \textit{t} &$K_{t}$ & Kalman gain at time \textit{t} \\ 
\hline
$f(\eta{L_{i,j}^{c}})$  & The execution result of the task $\eta{L}$ in the local & $\overrightarrow {sat}$ & The state vector of each sensor node\\ 
\hline
$E^{sleep}_{i}$ & Energy consumption of sensor node \textit{i} in sleep. & $E^{idle}_{i}$ & Energy consumption of sensor node \textit{i} in idle\\
\hline
$E^{check}_{i}$ & Energy consumption of sensor node \textit{i} in checking state. &$E^{work}_{i}$ & Energy consumption of sensor node \textit{i} in working state \\
\hline
$E^{com}_{i}$ & Communication consumption of sensor node \textit{i}. &${z}_{i,t}$ & the noise signal amplitude received at sensor \textit{i}. \\
\hline
$E^{trans}_{i}$ & Transmission consumption of sensor node \textit{i}. &$R_{i}$ & Received consumption of sensor node \textit{i}. \\
\hline
$T_{i}$ & Sending consumption of sensor node \textit{i}. &$E_{r}$ & The mean square tracking error for all sensor nodes.\\
 \hline 
 \end{tabular} 
\end{table*} 

The above studies are based on the assumption that edge servers have sufficient computing resource to process massive offloading tasks. However, it is not always feasible in practice due to the redundant and complicated data. Many mobile devices can share and provide their computation resource to edge servers, it is difficult to integrate the computing resource of mobile nodes and edge servers due to the dynamic network topology. An edge intelligence based dynamic resource management manner can be a flexible solution to meet the consecutive and accurate target tracking.

\section{System Model and Problem Formulation}
In this section, we propose a hierarchical network structure to facilitate the real-time target tracking. In this structure, the mobile node acts as the bridge for integrating sensing and computing resource, collecting sensing data with static nodes for collaborative computing with edge serves. Based on the consideration, a multi-objective optimization model is formulated to obtain the optimal target tracking performance.

\subsection{System Model}
The structure is shown in Fig. 1, there are two types of nodes: static (observation) sensor nodes and mobile nodes. For simplicity, we denote them by "SNs" and "MNs", respectively. Based on the functionality of network components, we divide the MTT system into three hierarchical levels: data sensing, mobile coverage and tracking, and intelligent computing and scheduling.  

As shown in Fig. 1, $\mathcal{M}$ heterogeneous SNs are deployed randomly to detect invading targets with diverse on-board sensors in a monitoring area, which is displayed at the bottom level (data sensing). Different from SNs, $\mathcal{N}$ MNs can process and compute the sensing tasks on the middle level (mobile coverage and tracking). During the task execution, MNs are not associated with any cluster, so that they can reduce blind sensing zones due to their flexible mobility. Considering the large-range data sensing, MNs can process the massive data by the proposed multi-resource data fusion algorithm to alleviate data transmission pressure. On the top level of intelligent computing and scheduling, scheduling decisions can be made by edge servers from the global viewpoint. Specifically, an intelligent cloudlet pattern is designed to alleviate computing pressure by the integration of resource of both MNs and edge servers. Based on the pattern, MNs can activate nearby SNs to observe targets in a real-time manner. The collaborative tracking scheme can ensure accurate tracking. Meanwhile, the tracking performance can also be saved in edge servers to conduct the following decision of node scheduling. For ease of reference, the key notations are summarized in the Table \uppercase\expandafter{\romannumeral1}.

We use $H(L,\tau_{ar})$ to represent an execution task, where $L$ is task size and $\tau_{ar}$ is execution deadline, i.e., the task is processed within $\tau_{ar}$. Once the task is executed, the execution time $t_\alpha \leq \tau_{ar}$ should be guaranteed. To address the problem, the task is portioned to $\eta L$ and $(1-\eta)L$, where $\eta L$ is implemented in mobile nodes and $(1-\eta)L$ is executed in edge servers. In this case, collaborative computing can reduce the computing latency to ensure time-sensitive requirements. We assume that time is discrete, and denote the time slot length and time slot index set by \textit{t} and $\mathcal{T}=\{0,1,2,...,\}$, respectively. The collected data is transmitted to mobile nodes or edge servers based the analysis of transmission model. MNs can process the data by the proposed multi-source data fusion algorithm. After that, all the data is located at edge servers, which implement the prediction of mobile target trajectory.   

\begin{figure*}[tp]
\captionsetup{justification=raggedright,singlelinecheck=false}
\centerline{\includegraphics[width=.7\linewidth]{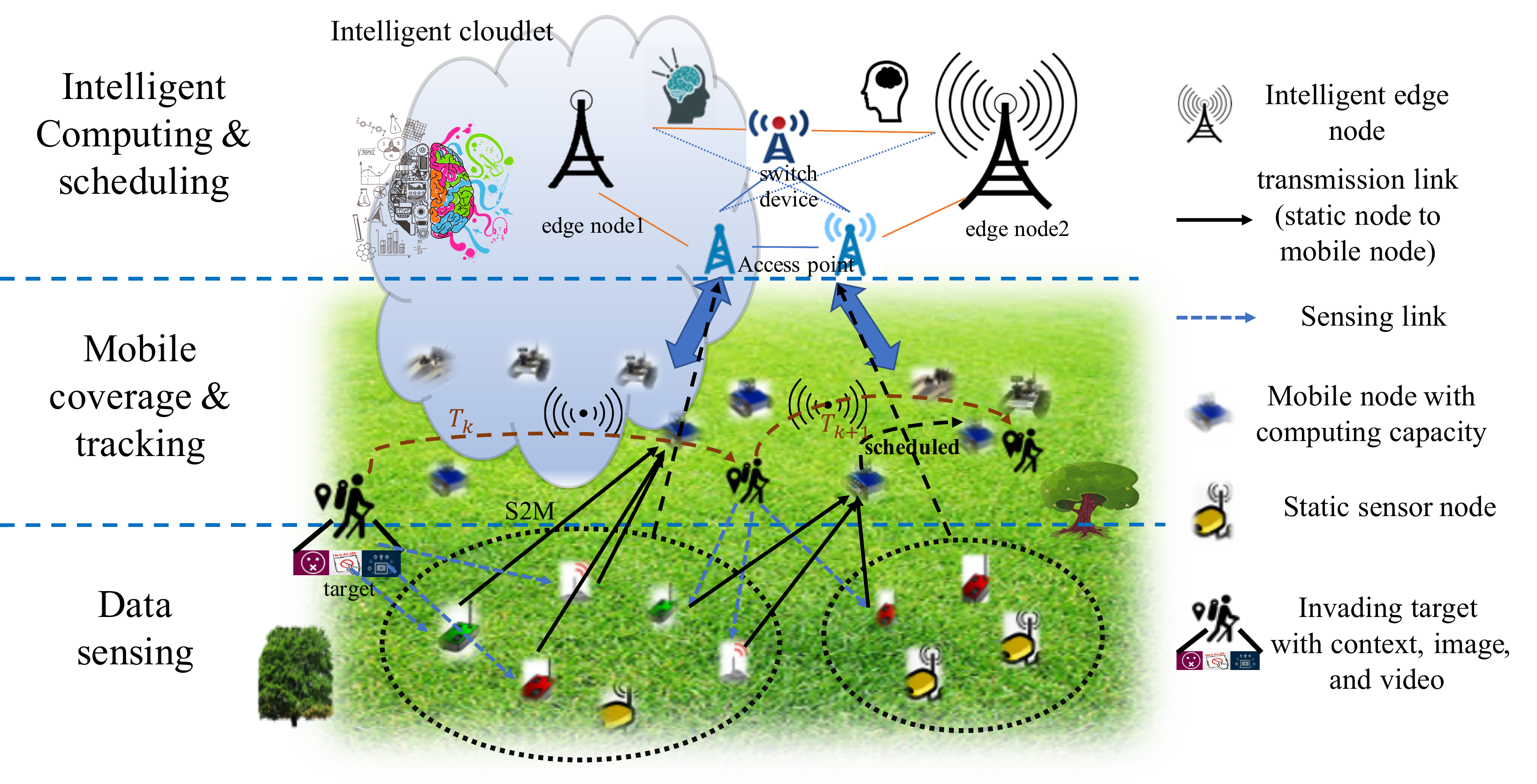}}
\caption{Illustration of a hierarchical target tracking structure. $M$ heterogeneous SNs are deployed randomly to detect invading targets with diverse on-board sensors on the bottom level. $N$ MNs can process and compute the sensing tasks on the middle level, so that blind sensing zones can be reduced based on their flexible mobility. Intelligent scheduling decisions are made by edge servers from a global view.}
\label{fig1}
\end{figure*}

\subsubsection{Analysis of Transmission Model}
To realize the real-time computing process, transmission model is formulated to optimize the offloading destination for low-latency transmission. The offloading destinations of sensing data include MNs and edge servers. The SNs can select their optimal destinations with the minimal bandwidth resource consumption. The  optional offloading space is represented as $\mathcal{A^{L}}=\{a_{l}\}=\{a_{i,j},a_{i,s}\}$. The SNs can select their optimal destinations collaboratively with the minimal bandwidth resource consumption. When radio bandwidth resource $a_l= a_{i,j}$ at the time slot $t$, the SN $i$ can offload sensing data to the MN $j$, otherwise, the sensing data can be offloaded to the edge server $s$. Assume that the SN $i$ has detected the invading target and $a_l= a_{i,j}$, sensing data can be transmitted to the neighboring MN $j$. If there exists multiple MNs, the SN $i$ can select the optimal destination node with the estimation of available radio bandwidth resource when there exists multiple MNs. If there no exists MNs within its communication range, the SN can select the optimal edge server to offload sensing data with $a_l=a_{i,s}$. 

For each sensor node, the transmission power and channel gain are denoted as $p^t$ and $g^t$ at each time slot \textit{t}, respectively. The transmission rate $r_{i,j}^{{S}\xrightarrow{}{}{M}}$ and $r_{i,s}^{{S}\xrightarrow{}{}{s}}$ are expressed as 
\begin{equation}
\left\{
\begin{split}
&r_{i,j}^{{S}\xrightarrow{}{}{M}}=a_{i,j} \times \log_2 (1+\dfrac{p_i^t\times g_i^t}{\delta_i}),\\
&r_{i,s}^{{S}\xrightarrow{}{}{s}}=a_{i,s} \times \log_2 (1+\dfrac{p_i^t\times g_i^t}{\delta_i}),
\end{split}
\right.
\end{equation}
where $\delta_i$ is system noise with Gaussian property. 

The corresponding transmission latency $t_{i,j}^{{S}\xrightarrow{}{}{M}}$ and $t_{i,s}^{{S}\xrightarrow{}{}{s}}$ are given by
\begin{equation}
\left\{
\begin{split}
&t_{i,j}^{{S}\xrightarrow{}{}{M}} = \dfrac{{L}}{r_{i,j}^{{S}\xrightarrow{}{}{M}}},\\
&t_{i,s}^{{S}\xrightarrow{}{}{s}} = \dfrac{{(1-\eta) L}}{r_{i,s}^{{S}\xrightarrow{}{}{s}}}.
\end{split}
\right.
\end{equation}

\subsubsection{Analysis of Data Fusion Model}
In order to track mobile targets with a high successful probability, a multi-element integration scheme is introduced to improve tracking accuracy. When MN \textit{j} involves tasks computing that is collected from nearby \textit{m} SNs. These values deviating from median ridiculously are removed. The updated values are represented as a set $\mathcal{F}$ and is given by
\begin{equation}
\begin{split}
\mathcal{F} &= \{f(\eta{L_{i,j}})\}, \forall i \in [1,m]\\
&=\mathop{max}\limits_{f(\eta{L}_{i,j})}|| \mathop{min}\limits_{f(\eta{L}_{i,j})}|| \dfrac{1}{m}\sum_{i=1}^m f(\eta{L}_{i,j})\\
& - f(\eta{L}_{i,j}) ||- f(\eta{L}_{i,j}) ||,
\end{split}
\end{equation}
where $f(\eta{L}_{i,j})$ denotes that the computing result occurred in the MN \textit{j}. The final performance set $\mathcal{P}$ is integrated by Cartesian operation with involved m SNs where $m \in \mathcal{M}$ and the MN \textit{j}, and is represented as 
\begin{equation}
\mathcal{P} = \mathcal{M} \times \mathcal{I} - \mathcal{F}.
\end{equation}

Intelligent scheduling is executed by computing the remaining task in edge servers. Nearby MNs and edge servers are made up of cloudlet for performing collaborative computing. The cooperative execution time $\tau_r$ is expressed as 
\begin{equation}
\tau_r = \dfrac{(1-\eta){L}}{f^e},
\end{equation}
where $f^e$ denotes intelligent computing capacity. Tracking time $\tau_{a}$ is consumed once mobile control is distributed with indicator $I_j^t=1$, where $I_j^t \triangleq \{0,1\}$. The system latency $t_\alpha$ is expressed as     
\begin{equation}
t_\alpha=\tau_c + \tau_\beta + I_j^t \times \tau_a \leq \tau_{ar},
\end{equation} 
where $\tau_c$ and $\tau_\beta$ denote the offloading latency and system computing time, respectively.

 \subsubsection{Trajectory Prediction Model}
The trajectory prediction, aiming at the minimum deviation, is modeled as Extend Kalman Filter (EKF) process, which incorporates prediction and update procedures \cite{Foder}. Unlike series forecasting or grey modeling methods, predicting mobility with inertial motion has excellent merit, especially for discrete-process control. The movement motion with respect to target assumed to be an acoustic is given by               
	\begin{equation}
	x_{t + 1|t} = F{x_t} + \omega_t,
	\end{equation}
where $x_t$ is location including $({x_t},{y_t})$, $\textit{F}$ is transfer matrix, and $\omega_t$ is noise matrix, namely Gaussian White noise \cite{Vodel}.

In the prediction process, the covariance matrix, i.e., $P_{t+1|t}=F\times P_{t}F^T$, is given to conduct prediction estimation. At the $\textit{t}$th time slot, the noise signal amplitude received at the sensor node $\textit{i}$ is repressed as $z_{i,t}= \sqrt{\dfrac{P_i^t}{1+(d_i^t)^{2}}}+\omega_{i,t}$, which is identity element of measurement vector $z_t$, where $d_{i,t}$ is the distance between the target and the $\textit{i}$th sensor node \cite{Braca,YSun,lkeda}.

In the update process, Kalman gain $K_t$ and deviation value $\widetilde{y}$ is acquired for consecutive prediction process. The measurement residual, i.e., $\widetilde{y} = z_t - \int{H_t}$, estimates prediction process, where $H_t$ is the measurement matrix mapping the actual state space into the measurement space. Kalman gain considering minimum mean square error as objective function is derived as 
\begin{equation}
	{K_{t + 1}} = {P_{t + 1|t}}H_{t + 1}^TS_{t + 1}^{ - 1},
	\end{equation}
 where ${S_{t + 1}} = {H_{t + 1}}{P_{t + 1|t}}H_{t + 1}^T$ is the innovation covariance. The covariance matrix updated by iteration process is expressed as ${P_{t + 1|t + 1}} = (I - {K_{t + 1}}{H_{t + 1}}){P_{t + 1|t}}$ \cite{Hu}. Consequently, the updated state estimation is given by
 \begin{equation}
 x_{t+1|t+1}=x_{t+1|t} + K_t \times \widetilde{y}.
\end{equation}         
	
\subsection{Problem Formulation }
Execution cost is one of key measures for target tracking performance and it is invoked to optimize scheduling strategy for the MTT-WSN. Thus, self-states of sensor nodes are divided into four categories which are transferred with each other for saving energy. These states, including sleep, check, idle, and work, are represented as a vector $\overrightarrow {sat} = [{s^{sleep}},{s^{check}},{s^{idle}},{s^{work}}]$. Sensor nodes stay sleep state when there are no tracking tasks. Then, nodes are activated to idle states for detecting targets. When collecting sensing data, states of nodes are changed into checking state to execute targets position. Mobile nodes are scheduled to track targets cooperatively once work states are enabled. As shown in Fig. 2, indicator vector $[1,0,0,0]$ signifies that the sensor maintains sleep state.

\begin{figure}[!t]
\captionsetup{justification=raggedright,singlelinecheck=false}
\centerline{\includegraphics[width=.75\columnwidth]{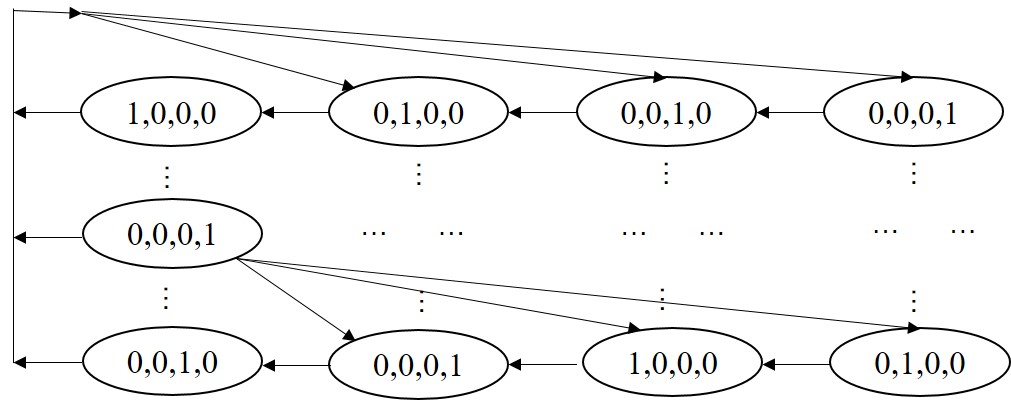}}
\caption{All the cases of the switched states.}
\label{fig2}
\end{figure}

To be simplify, $\textit{P}_0$ denotes the energy cost of sleep state at a unit time \textit{t}. During a tracking period $\gamma$, the energy cost of sensor \textit{i} is given by
\begin{equation}
	E^{sleep}_{i}=\int_{0}^{\gamma} {{\textit{P}_0}dt},
	\end{equation}

The energy cost of idle states is obviously more than that of sleep states. It is assumed that $\textit{P}_{idle} = {\eta _w}{P_0}$ is the unit energy cost of idle states where ${\eta_w}>1$. Energy consumption during a tracking period is given by
 	\begin{equation}
	E^{idle}_{i}=\int_{0}^{\gamma} {{\textit{P}_{idle}}dt} ,\forall t \in \mathcal{T}
	\end{equation}

The scheduling schemes among sensors are same and independent. Assume that the sensor node \textit{i} is scheduled with the probability $\phi_i$, periodic consumed energy of sensor \textit{i} staying the check state is represented as $E_i^{check} = \frac{{\phi _i - {{(\phi _i)}^{k + 1}}}}{{1 - \phi _i}}\int_{0}^{T} {{\eta _w}{P_0}dt}$. when $\textit{k}=1$, the energy consumption of the sensor \textit{i} can be represented as
	\begin{equation}
	E_i^{check} = E_i^{trans} + E_i^{com} + \phi _i\int_0^{\gamma} {{\eta _w}{P_0}dt} ,
	\end{equation} 
where $E_i^{trans} = 2\times {\varepsilon _{elec}} \times ({q_t} + {q_s}) + {\varepsilon _{amp}} \times {q_r} \times {d^2} $, and $\varepsilon _{elec}$ and $\varepsilon _{amp}$ are circuit and gain consumption of amplifier consumption for transmitting 1 bit, respectively. $q_t$, $q_s$, and $q_r$ denote the different data sizes, respectively. $E_i^{com} = \kappa aL{f^2}$, where $\kappa$ is the effective switched capacitance depending on the chip architecture. $a$ is a real number limited in the [0,1].

The scheduling of sensor nodes is depended on the physical distance between sensor nodes and invading targets as well as their self-energy, which is defined as 
\begin{equation}
{\textit{con}_{i,t}} = {\omega _1}{R_{i,t}} + {\omega _2}e^{-d_{i,t}},  
\end{equation}
	where ${\omega _1}$ and ${\omega _2}$ are weight coefficients satisfying ${\omega _1} + {\omega _2} = 1$. ${R_{i,t}}$ and $d_{i,t}$ is self-energy of sensor node \textit{i} and the distance to targets. It is noteworthy that$Con_{i,t}$ is normalized.

When MNs are scheduled, the mobile energy cost is represented as
\begin{equation}
	E_i^{work} = v\gamma\varpi_v ,
	\end{equation}
where $\varpi_v $ is per unit energy cost. 	
	
Consequently, different energy cost is represented as a vector $\overrightarrow{E}^{state}_{i}$, i.e., $\overrightarrow {E}^{state}_{i}  = [{E^{idle}},{E^{sleep}},{E^{check}},{E^{work}}]$. When \textit{m} nodes are deployed in a monitoring area including sensor nodes and mobile nodes. The long-term execution cost is formulated as 
\begin{align}
	&\qquad P1:\mathop {\min } \limits_{{\phi _i},{\vec{sat}}} \,\,\frac{1}{T} {{{\sum\nolimits_{\gamma = 1}^{T} \sum\nolimits_{i = 1}^{m} {I^{R_{4\times1}}E_i^{state}} }}} \tag{15}\\
	&s.t.\quad
	\begin{cases}
	C1: t_\alpha \le \tau_{ar}, \notag \\ 
	C2:E_{i,j}^{state} \le {E_{i,\max }},\,j \in \{ 1,2,3,4\} \notag \\
	C3:\sum {\mathop {\arg }\limits_i (co{n_i} > {\phi _{i}^{\min }})}  \ge \sum {\mathop {\arg }\limits_i {E_r}}, \notag \\
	C4:0 \le co{n_i} \le 1.\notag\\
	C5:I_{k,j} \in \{0,1\}, and \sum_{j=1}^{4}=1. \notag
	\end{cases}
	\end{align}

\textit{C1} denotes that tasks can be completed within deadline. \textit{C2} is power control constraints, i.e., execution power cannot exceed maximum power. \textit{C3} is the tracking accuracy constraint, i.e., ${E_r} = \sqrt {{{(\frac{{\sum\nolimits_{i = 1}^m {{x_i}} }}{m} - {x_t})}^2} + {{(\frac{{\sum\nolimits_{i = 1}^m {{y_i}} }}{m} - {y_t})}^2}}$ is required accuracy \cite{Mahfouz}. \textit{C4} indicates that the normalized tracking capacity is limited in a feasible range. \textit{C5} denotes that only one state is existed at each time slot for each sensor.

\begin{figure*}[tp]
\captionsetup{justification=raggedright,singlelinecheck=false}
\centerline{\includegraphics[width=.7\linewidth,]{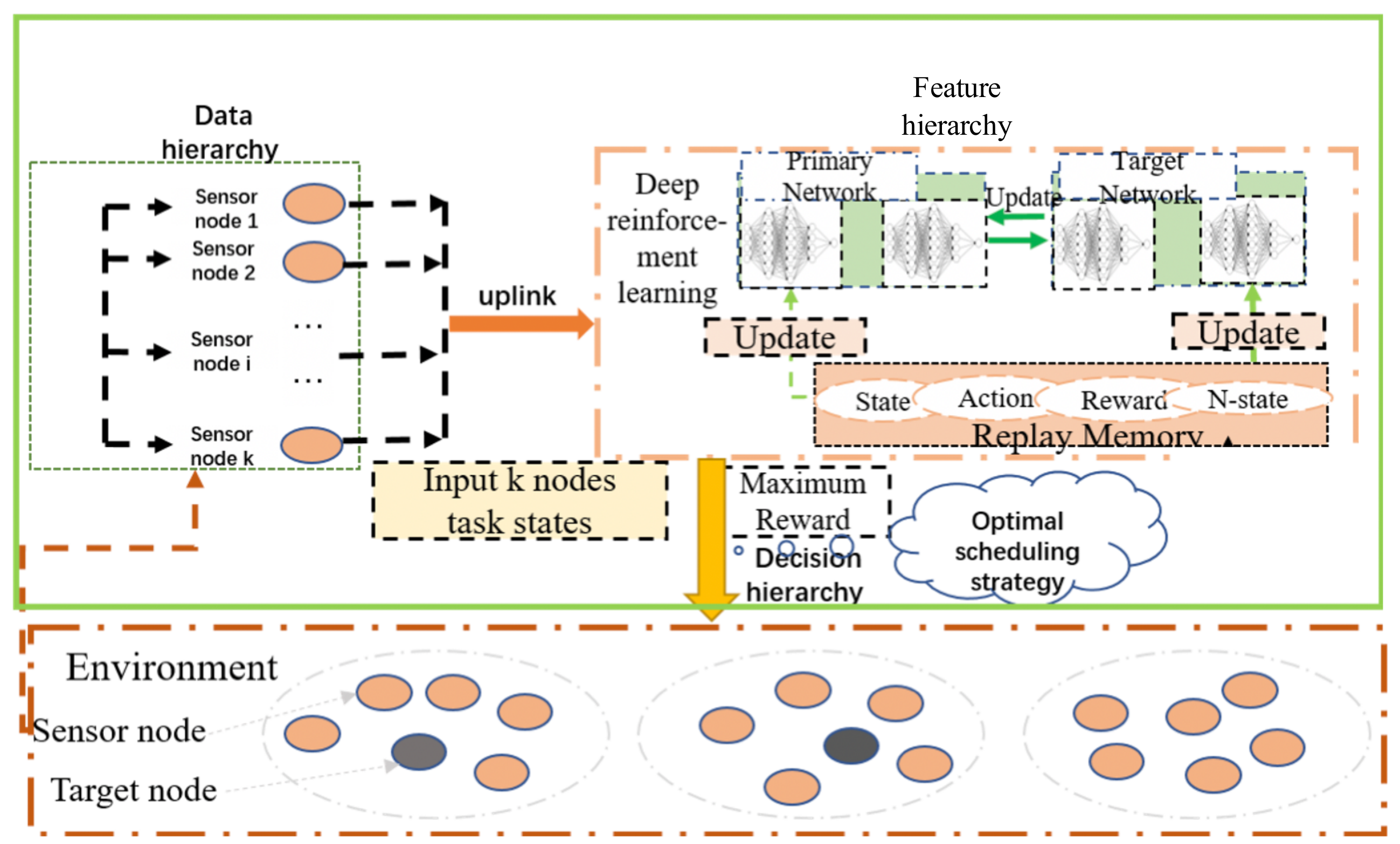}}
\caption{The long-term dynamic resource allocation algorithm. state information in the data layer is collaboratively swapped and collected from the MTT environment. Data is transmitted to the feature layer for mobile trajectory prediction. The data is trained using the prediction neural network architecture. The prediction results are transmitted to the decision layer that makes node scheduling.}
\label{fig2}
\end{figure*}  

\section{Long Term Dynamic Resource\\ Allocation Algorithm}
To acquire the optimal target tracking strategy, we propose a long-term dynamic resource allocation algorithm. In this algorithm, computing and tracking decisions are executed synchronously for the time-sensitive MTT-WSN requirements.

\subsection{The Markov Decision Process}	
In the MTT network, system actions are only depended on the current system states during tracking process and problem \textit{P}1 is regarded as a long-term optimal average system cost process. Consequently, the \textit{P}1 is formulated as an MDP model incorporating state space, action space, reward formulation, as well as state transferring equation.    

	$\bullet$ \textit{\textbf{The state space:}} At the \textit{t}-th time slot, the state space includes trajectory prediction, tracking capacity (namely the dump energy, the distance between the target and sensor nodes), and node states. Thus, the state is expressed as $s(t) = \big \{ x(t), K_{t}, con(t), t_{\alpha}, E^{sleep}_{i,t}, E^{idle}_{i,t}, E^{check}_{i,t}, E^{work}_{i,t}, E^{trans}_{i,t}, \\E^{com}_{i,t} \big \}$.

	$\bullet$ \textit{\textbf{The action space:}} The action space is related to the tracking performance, which results from deep training and learning as well as the next state. The specific action space is represented by $a(t)=\big \{a_{l}, \beta_{i} \big \}$, where $\beta_{i} \in \{0,1\}$ indicates that the sensor \textit{i} is scheduled or not.
	
	$\bullet$ \textit{\textbf{The reward formulation:}} The action experience is conducted by the reward to encourage the better performance. The causal reward and action are coupled by the reward formulation $r(t) = {k_1}e + {k_2}a + {k_3}q$, where $e$, $a$, and $q$ denote the system energy consumption, mean square error, and punishment for unsatisfied performance, respectively, and $k_1$, $k_2$, and $k_3$ are the corresponding weight coefficients and satisfy $k_1 + k_2 + k_3 =1$. 
	
	\begin{algorithm}
		\renewcommand{\algorithmicrequire}{\textbf{Input:}}
		\renewcommand{\algorithmicensure}{\textbf{Output:}}
		\caption{ The Long-Term Dynamic Resource Allocation algorithm}
		\begin{algorithmic}[1]
			\REQUIRE ~ 
			$Q(\theta)$, iteration number $\textit{N}$, discount factor $\gamma$, gradient descent rate $\eta$, the number of step $\mathcal{W}$, the transfer matrix, the measurement matrix, round = 1000, and count=0;
			\STATE Initialize Q-network with weights $\theta$, action-cost function $\textit{Q}$, experience replay memory, and $\mathcal{F}=\mathcal{M}$;
			\FOR {each time slot $t \in T$}
			\FOR {each sensor $i \in \mathcal{F}$}
			\STATE Predict the target trajectory based on Eq. (9); 	
			\STATE Update the target position;
			\STATE Store to a temporary array (cache).
			\ENDFOR
			\ENDFOR
			\STATE Return the caching array.
			\FOR {each sensor node $k \in \mathcal{M}$}
			\STATE $\mathcal{F} = \mathcal{F} \backslash \mathop{argmax}\limits_{i} (f(\eta{L_{i,j}}))$ based on Eq. (3);
			\ENDFOR
			\STATE Return $\mathcal{F}$.
			\WHILE {count $\le$ round}	
			\STATE Train the data of data layer;
			\STATE Update network parameters and weights;
			\STATE count = count +1;
			\ENDWHILE
			\STATE Acquire output values;
			\STATE Storage output values into the replay memory;
			\FOR{each interaction with environment $episode \in \mathcal{W}$}
			\FOR{each time $t \in \textit{T}$}
			\STATE Select action $a$ and the corresponding state randomly;
			\STATE The corresponding reward $r$ is computed and stored;
			\STATE Compute gradient function based on Eq. (19);
			\STATE Calculate the gradient of weight $\theta$;
			\STATE Updating: $Q(s,a) \leftarrow R(s,a) + \gamma \mathop {\max }\limits_a Q(s,a)$s;
			\IF {\textit{step} == \textit{N}}
			\STATE Reset $Q_\theta$;
			\ENDIF
			\ENDFOR
			\STATE  This episode is terminated.
			\ENDFOR 
			\ENSURE ~ 
			The optimal scheduling cost;
		\end{algorithmic}
	\end{algorithm}

	$\bullet$ \textit{\textbf{The state transforming equation:}} The system interaction as an important step is revealed to obtain the optimal system benefits. The probability model is formulated based on the Markov chain, in which the sample values stored in the memory are based on the primary and forward data. The equation is expressed as $P({s^{'}}|(s,a)) = P({s^{'}}(p(t),con(t),E^{state}_{i,t})|s(x(t), K_{t}, con(t), E^{sleep}_{i,t}, E^{idle}_{i,t},\\ E^{check}_{i,t}, E^{work}_{i,t}, E^{trans}_{i,t}, E^{com}_{i,t}),a_{l,t}, \beta (t), E^{state}_{i,t}, E_{r})$.			

\subsection{Analysis of long-term dynamic resource allocation algorithm}	
As shown in Fig. 3, the proposed LTDRA integrates the hierarchies of data, feature and decision. Specifically, in the data layer, state information, including target mobility and own available capacity, is collaboratively swapped and collected from the MTT environment. Data is transmitted to the feature layer for mobile trajectory prediction. The data is trained in the prediction neural network architecture, which can analyze mobile trajectory in the following time slots (i.e., prospective mobile trajectory). The prediction results are transmitted to the decision layer that makes the strategy of node scheduling, based on the execution deadline and system energy. In the decision layer, smart agent implements self-driven learning by interacting with environment, and the agent iterates node scheduling strategy by sampling from the updated replay memory. A Deep Reinforcement Learning (DRL) algorithm is proposed to overcome data correlation by sampling from the replay memory. Besides, the prediction data is fed into the replay memory of decision layer to acquire fresh knowledge. The prediction process is represented as
$\left[ \begin{array}{l}
x(k + 1)\\
x(k + 2)\\
\quad \;\; \vdots 
\end{array} \right] = \left[ \begin{array}{l}
x(1),x(2),\; \cdots ,\;x(k)\\
x(2),x(3),\; \cdots ,\;x(k + 1)\\
\quad  \vdots \quad \;\; \vdots \quad \; \ddots ,\;\quad \; \vdots 
\end{array} \right]\left[ \begin{array}{l}
{\chi _1}\\
{\chi _2}\\
\;{\kern 1pt} {\kern 1pt}  \vdots \\
{\chi _m}
\end{array} \right]$, where $\chi_{i}$, $i \le m$ is the estimation coefficient.

In the decision layer, smart agent implements self-driven learning by interacting with environment and iterates node scheduling strategy by sampling from the updated replay memory. State set forming state space is fed into primary network. The node scheduling actions are acquired from the target network. The two neural networks are synchronously executed to facilitate respective learning process. The target network can evaluate the current state-action pair by using cost function, i.e., $Q(s_t,a_t,\theta)$. The Q-value that is a unity value can replace the multi-object optimization model for the optimal node scheduling.

The cost function is derived by Bellman equation. The \textit{P}2 is accumulated reward expectation and represented as
	\begin{equation}
	P2:R(s) = \frac{1}{T}E\{ \sum\nolimits_{t = 0}^T {{\gamma ^t}r(s(t),a(t))|s(0) = 0} \}, \tag{16}\notag
	\end{equation}  
	where $\gamma$ is the discount factor. $P2$ denotes the average exception of acquired reward, and $E{~}$ is the exception for a long-term cumulative process. The iteration process is given by
	\begin{equation}
	{R^{*} }(s) = \mathop {\min }\limits_{a \in A} \{ c(s,a) + \gamma \sum\limits_{s^{'} \in S} {P(s^{'}|(s,a)){R^ * }(s^{'})} \}, \tag{17}\notag
	\end{equation}
where $c(s,a)$ is the reward under the condition of state \textit{s} and action \textit{a}, the optimal scheduled strategy is acquired by
	\begin{equation}
	{\pi^{*} }(s) = \mathop {\arg \min }\limits_{a \in A} {R^{*} }(s). \tag{18}
	\end{equation}
	
Unfortunately, the problem \textit{P}2 is only proper for those data with low dimensional character. Reducing dimension may result in the high computation complexity, which is not feasible in the MTT network with time-sensitive character. However, an alternative method can obtain an approximate solution to replace the problem \textit{P}2 and can meet acceptable time complexity and space complexity. The approximation approach is formulated and given by
	\begin{equation}
	Q(s,a) = c(s,a) + \gamma \mathop {\min }\limits_{a^{'} \in A} R(s^{'}), \tag{19}
	\end{equation}
the corresponding updating process is $Q(s,a) = (1 - \eta )Q(s,a) + \eta (c(s,a) + \gamma \mathop {\min }\limits_{a \in A} Q(s,a))$, where $\eta$ is learning rate.

In the LTDRA algorithm, Q-reality and Q-estimate, i.e., $Q_{\theta^{'}}$ and $Q_\theta$, are formulated to present the primary network and the target network, respectively. The subscripts $\theta$ and $\theta^{'}$ are updated after each iteration and the reward as output is obtained to evaluate each action. The optimal scheduling can be reaped by maximizing each reward. Algorithm flow is specifically reflected in the Algorithm 1.

\subsection{Analysis of Algorithm Complexity}	
In the target tracking network, there exists $\mathcal{M}$ sensor nodes. We analyze the deep reinforcement learning algorithm from a macroscopic viewpoint. the time complexity is $\mathcal{O(WT|M|)}$. Consider the internal algorithm flow, we obverse that one action is randomly sampled from a list of actions. Thus, it has a time complexity $\mathcal{O} (1)$ in the per iteration. In the primary network, the complexity of matrix inversion is $\mathcal{O}(k(\theta))$, where $k(\theta)$ is a function of the depth and number of the hidden layers $\theta$. Finally, the whole network time complexity is represented as $\mathcal{O}(k(\theta)\mathcal{WT|M|})$. Although we add the prediction neural network, the time complexity is not increased, which is discussed in Section V.

\begin{table}[htbp] 
 \caption{\label{tab:test}The simulation parameters of MTT-WSN system} 
 \centering
 \begin{tabular}{|l|l|l|} 
  \hline 
   Parameter description&  Value  \\ 
  \hline 
  The invading range & $200m \times 200m$ \\ 
  \hline
  The number of target nodes&  1  \\ 
  \hline
  The number of sensor nodes&  56  \\ 
  \hline
  The number of mobile nodes & 6\\
  \hline
  The tracking velocity & 1m/s\\
  \hline
  The total energy for each sensor node & 40\textit{J}\\
  \hline
  The initial coordinate of target node & (0,50m)\\
  \hline
  The energy consumption with the sleep mode & 0.1\textit{J}/unit time\\
  \hline
  The energy consumption with the idle mode & 0.2\textit{J}/unit time\\
  \hline
  The energy consumption with the check mode & 0.6\textit{J}/unit time\\
  \hline
  The energy consumption with the work mode & 1.5\textit{J}/unit time\\
  \hline
  The tuning range learning rate & [0.1, 0.9]\\
  \hline
  The discount factor of reward function & 0.9\\
  \hline
  The size of min-batch & 32\\
  \hline
  The size of replay memory & 500\\
  \hline
 \end{tabular} 
\end{table}	
	
\section{Performance Evaluation}
Extensive simulation experiments are conducted to evaluate the proposed LTDRA algorithm based on our gathered multitude simulations, including tracking accuracy and system energy consumption. 
\subsection{Simulation Setup}
we use the version of Python 3.7 and the TensorFlow architecture to build the MTT scenario. The main influencing factors, including system energy consumption, tracking accuracy, and execution latency, have been programmed and evaluated to verify the efficiency of our proposed algorithm. Specifically, we design a square monitoring area where sensor nodes are deployed randomly in the initial stage. As shown in Figure 4, those red and blue solid circles in the area denote MNs and SNs, respectively. The number of MNs is set as 6, and that of SNs is set as 50. The noise covariance is set as $\delta_x$ = $\delta_y$ = 1, and $\textit{N}$ $\sim$ $(0,1)$. The total energy of each static sensor node is 40J. The initial location and velocity of the mobile target are set among the range $(0m,150m)$, and $(0 m/s,1m/s)$, respectively. The initial invading direction can be also arbitrary. To clearly express the simulation parameters, we summarize the important parameters in Table \uppercase\expandafter{\romannumeral2}.
For comparison, we introduce four benchmark strategies.
\begin{figure}[!t]
\captionsetup{justification=raggedright,singlelinecheck=false}
\centerline{\includegraphics[width=.75\columnwidth,]{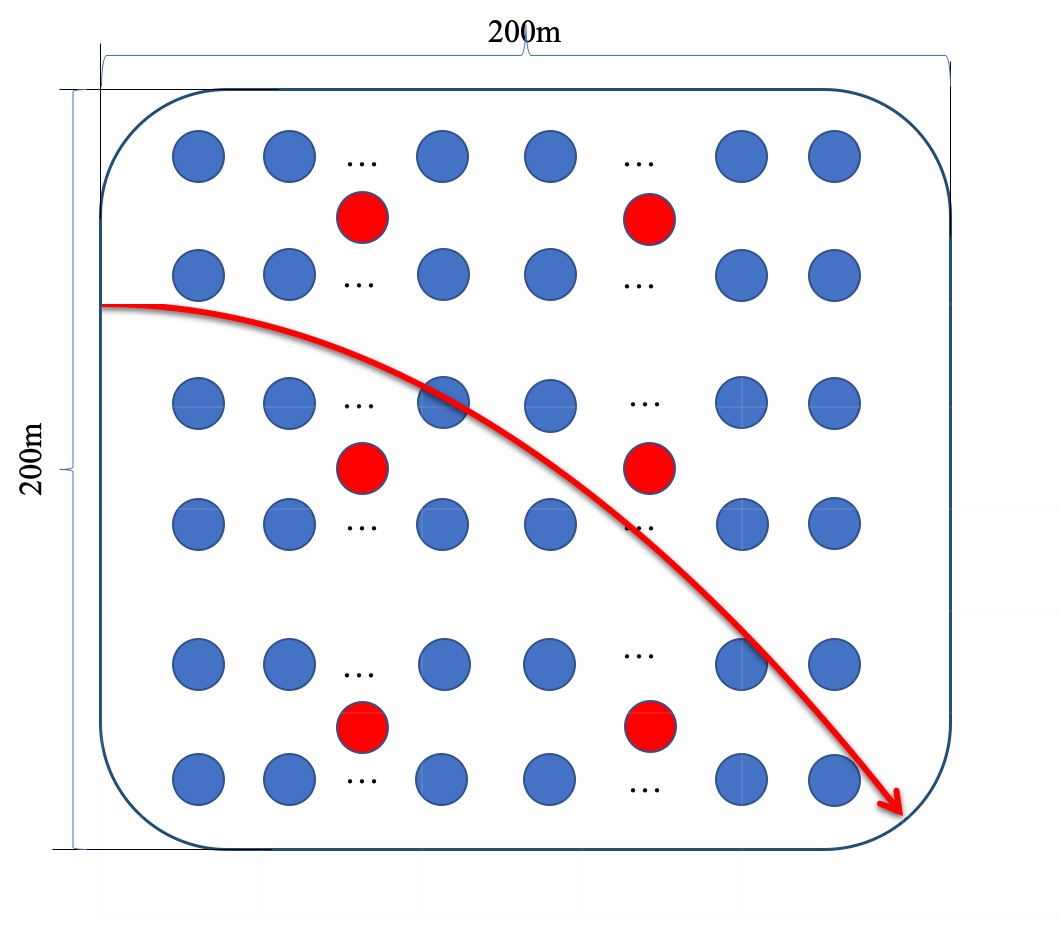}}
\caption{The random trajectory of mobile target.}
\label{fig5}
\end{figure}
\begin{figure}[!t]
\captionsetup{justification=raggedright,singlelinecheck=false}
\centerline{\includegraphics[width=.75\columnwidth,]{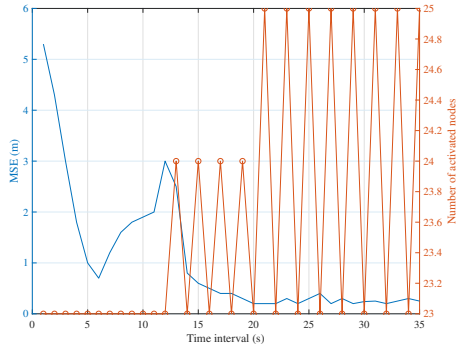}}
\caption{The MSE of target tracking versus the number of activated sensors.}
\label{fig6}
\end{figure}

\begin{figure}[htbp]
\captionsetup{justification=raggedright,singlelinecheck=false}
\centering 
\subfigure[]{
\includegraphics[width=.75\columnwidth,]{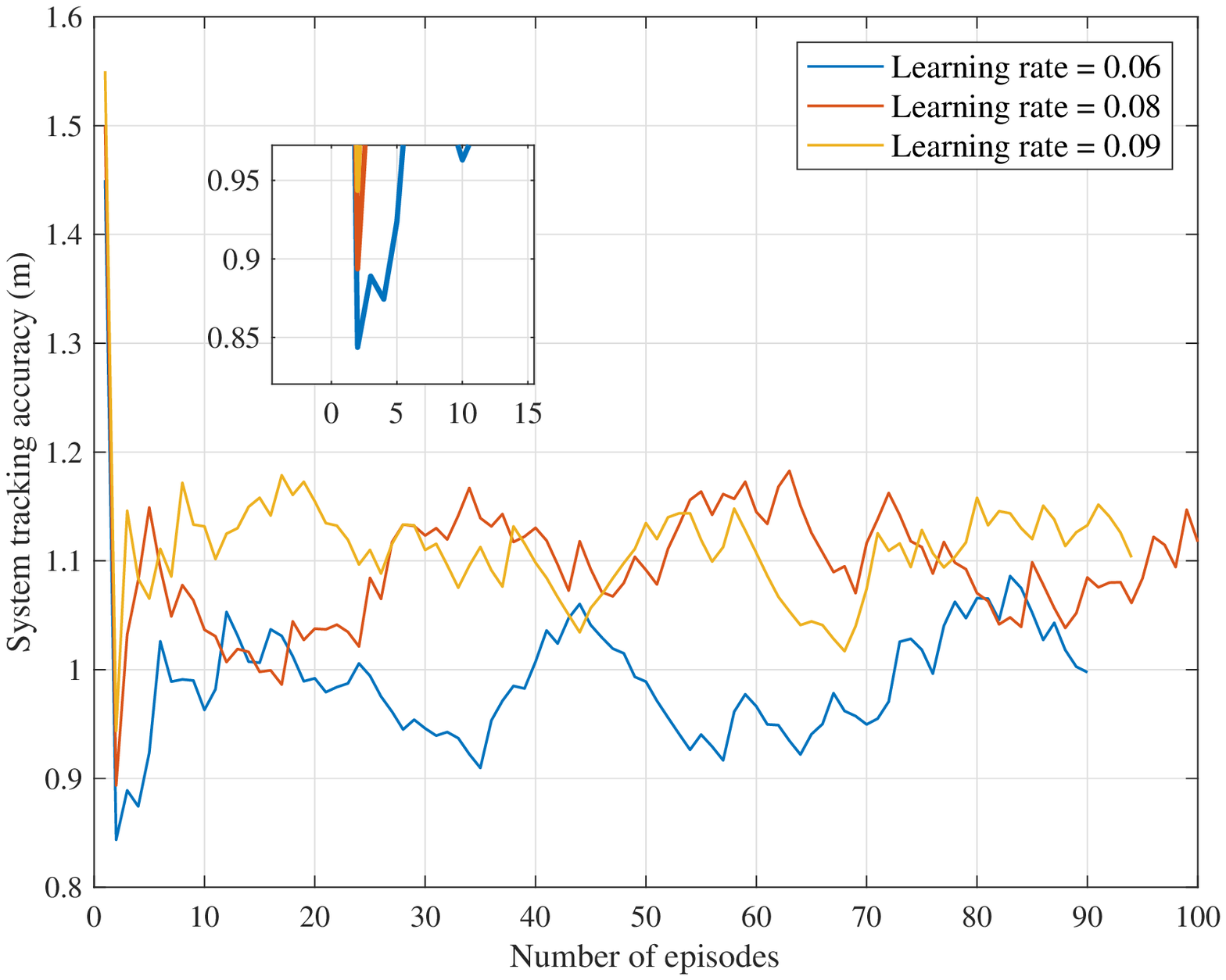}
}
\quad
\subfigure[]{
\includegraphics[width=.75\columnwidth,]{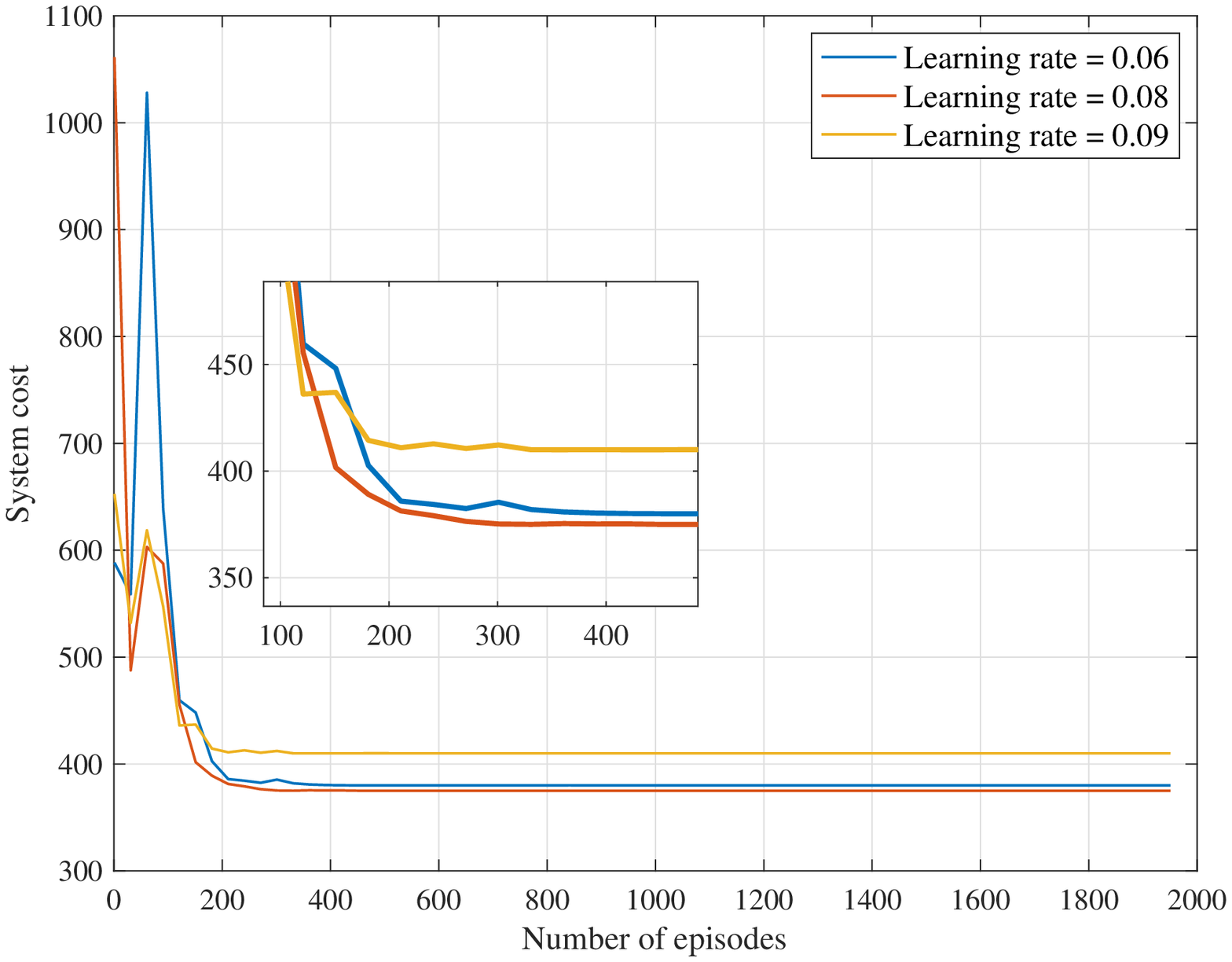}
}
\caption{Learning performance of the tracking accuracy (a) and system cost (b), where the proportion of the past data and the prospective data is 2:1.}
\label{fig9}
\end{figure}

\begin{figure}[htbp]
\captionsetup{justification=raggedright,singlelinecheck=false}
\centering 
\subfigure[]{
\includegraphics[width=.75\columnwidth,]{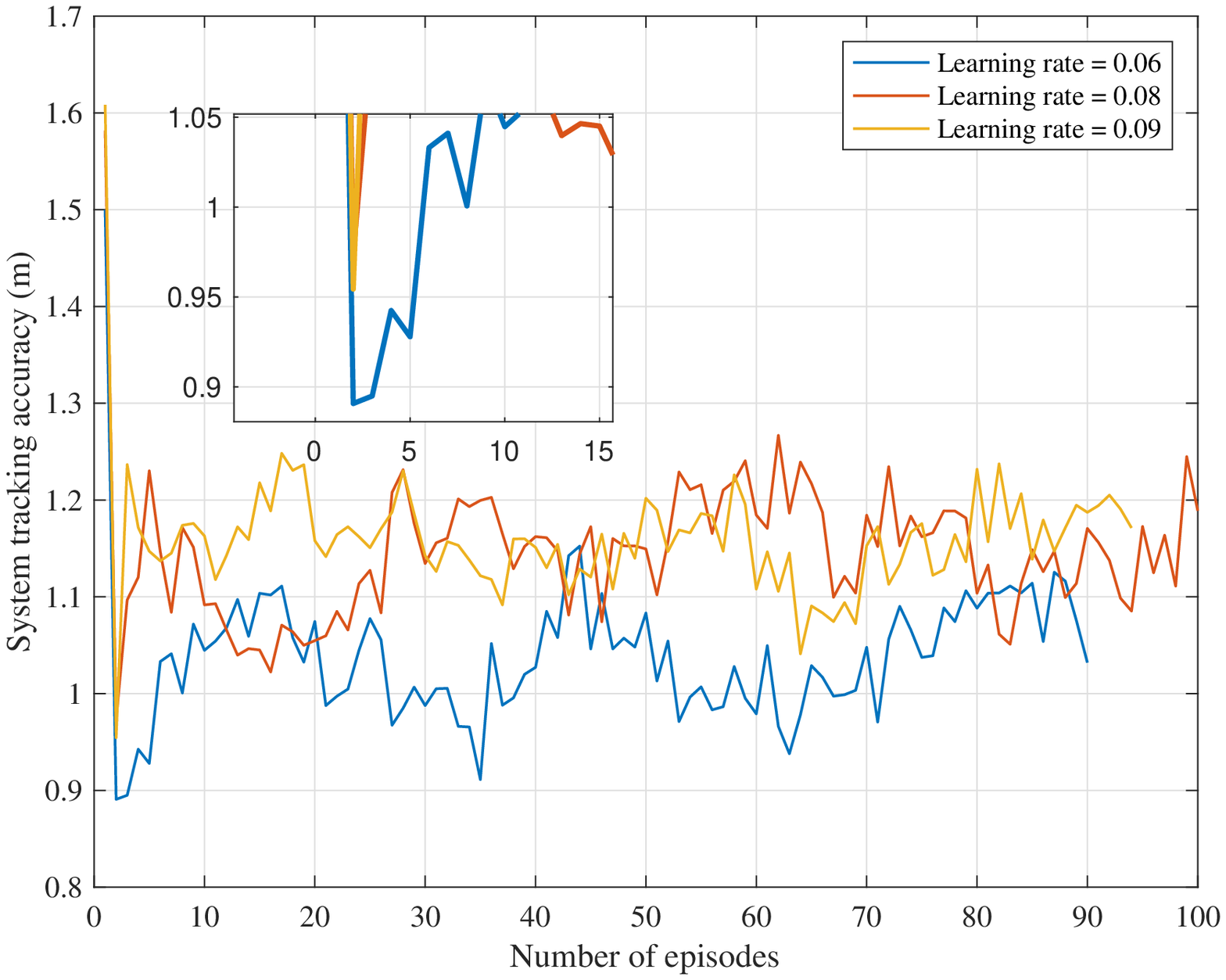}
}
\quad
\subfigure[]{
\includegraphics[width=.75\columnwidth,]{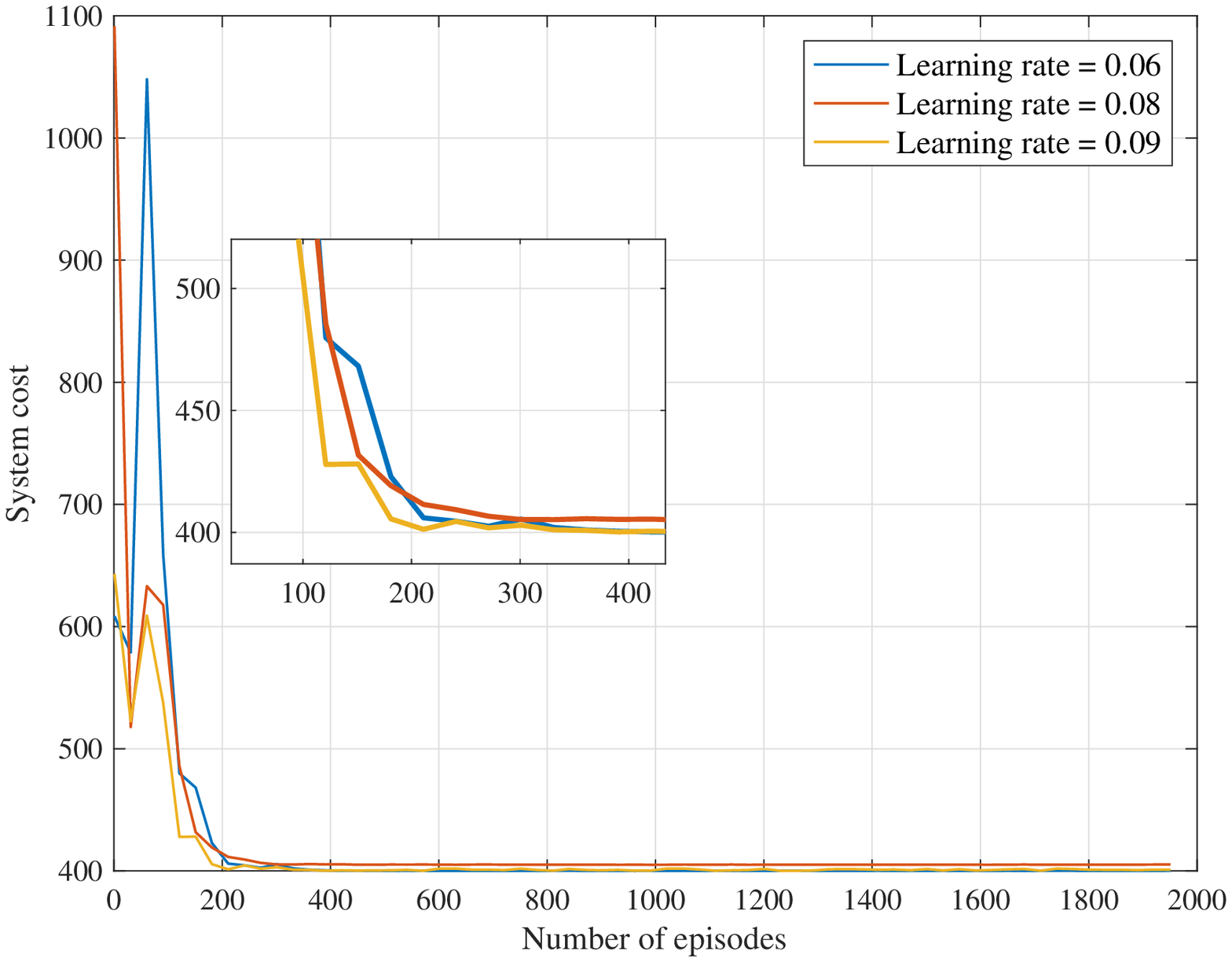}
}
\caption{Learning performance for tracking accuracy (a) and system cost (b), where the proportion of the past data and the prospective data is 3:1.}
\label{fig9}
\end{figure} 

\begin{itemize}
\item[•] \textit{Non-cooperative Scheme}: This scheme adopts DRL framework and prediction network. The collaborative between mobile nodes and the edge servers is not considered. 
\item[•] \textit{Deep Q-learning Scheme}: This scheme barely provides the DRL neural network without prediction network, and collaborative scheme is incorporated.
\item[•] \textit{Greedy Scheme}: This scheme with deep learning architecture aims to the execution cost minimization and selects sensor nodes with sufficient energy at each time slot \textit{t}.
\item[•] \textit{Random Selection Scheme}: When task is executed in an edge server, the selected probability of each sensor is limited in range [0,1] with the aided of the DRL architecture. The generated values are resorted by descending order, sensors are scheduled if the corresponding probability is greater than 0.5.  
\end{itemize}
	
\subsection{Results Discussion}
In this subsection, we verify the feasibility of prediction scheme, show the efficiency of proposed algorithm, and demonstrate the impacts of variable system parameters, respectively.

In Fig. 5, the Mean Squared Error (MSE), i.e., $MSE(t)\mathop  = \limits^\Delta  \frac{1}{N}{\sum\nolimits_{i = 1}^N {{{({x_i}(t) - {x_t}(t))}^2} + ({y_i}(t) - {y_t}(t))} ^2}$, is formulated to evaluate the prediction accuracy. The algorithm is terminated while the upper bound of round is meet. Fig. 5 depicts the tendency of MSE and number of activated sensors. The value of MSE is reduced gradually then remains stable. When $\textit{t}=13$, the number of activated sensor nodes begins to increase and position accuracy of the nodes also is improved. Besides, the position accuracy is convergent gradually although the number of the activated sensor nodes is undulant in a certain range. The reason why the trend is undulant is that the instability of the EKF and the prediction neural network. Specifically, fitting errors may exist when nonlinear motion is matched into linear motion which is processed in the EKF. Besides, the training process may also cause a fluctuation in the neural network.  

We show the effectiveness of target tracking and demonstrate the impact of different memory ratios of historical data size and prospective data size on prediction accuracy and system consumption. First, system performance is illustrated in Fig. 6 and Fig. 7 which includes prediction accuracy and system energy consumption under different memory ratios. It is noted that the mentioned tracking accuracy is equivalent to the mentioned trajectory prediction accuracy. When the ratio is 2:1, as shown in Fig. 6 (a), the prediction error decreases dramatically in the early stage. After that, the tracking error approaches a stable status which oscillates intensely in a certain range of [0.8, 1.2]. Fig. 6 (b) provides the tendency of system energy consumption. We can observe that the system energy consumption can decrease smoothly and tend to a steady state when the iteration reaches 500 approximately. When the ratio is 3:1, as shown in Fig. 7 (a), the convergent speed in prediction accuracy is slower than that of Fig. 6 (a). In Fig. 7 (b), the number of the iteration is the same as that of Fig. 6 (b) but the convergent energy consumption is higher. 

The following observations can be found in Fig. 6 and Fig. 7. Firstly, the range of prediction error is limited to $[0.8, 1.2]$, which is accepted in many practical tracking scenarios. The trends of system energy consumption and prediction accuracy are convergent and stable as the number of the iteration increases. This implies that our proposed algorithm can ensure constant tracking performance. Secondly, the convergent speed exists slightly different. The reason is different learning rates influence steps of gradient descent, which can generate different weight values in the training process. Finally, it can be observed that our proposed algorithm can achieve a quick convergence with different learning rates based on the combination of prediction neural network and reinforcement learning.

	 Fig. 8 makes the comparison among different scheduling strategies on system cost. With the increasing number of the iteration, all the scheduling schemes can achieve their own goal to reduce system energy consumption. The following observations are found from this figure. Firstly, the random selection scheme performs the highest system energy consumption. The main consumption is generated due to the mobility. Secondly, the proposed LTDRA algorithm obviously reduces the system consumption compared with the other four benchmarks. In the numerical analysis, the proposed algorithm reduces 14.5\%, 31.6\%, 42.8\% and 47.4\% approximately in system energy consumption compared with the deep Q-learning scheme, non-cooperative scheme, greedy scheme and random selection scheme.  

Fig. 9 provides the comparison of the tracking accuracy based on different scheduling schemes. In order to improve the system prediction accuracy, edge servers and mobile nodes acquire the whole system status to discover the optimal scheduling scheme cooperatively. For the greedy scheme, it always seeks these sensor nodes with sufficient energy while the accurate prediction cannot be guaranteed. For the non-cooperative scheme, mobile nodes only collect and transmit sensing data to the edge server. In this case, computing results may be high-latency due to massive data transmission. In contrast, the random selection scheme performs worst since the number of scheduling nodes is random. Compared with the non-cooperative scheme, the deep Q-learning algorithm outperforms in system energy consumption. This implies that the collaboration computing is of significance. The proposed scheme performs the lowest system energy consumption based on the joint optimization of tracking accuracy and system energy consumption with the coupled architecture including the deep reinforcement learning and the prediction network.
 \begin{figure}[!t]
\captionsetup{justification=raggedright,singlelinecheck=false}
\centerline{\includegraphics[width=.75\columnwidth,]{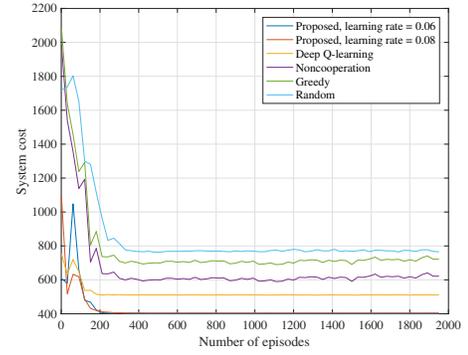}}
\caption{System energy consumption versus number of iteration.}
\label{fig10}
\end{figure}

\begin{figure}[!t]
\captionsetup{justification=raggedright,singlelinecheck=false}
\centerline{\includegraphics[width=.75\columnwidth,]{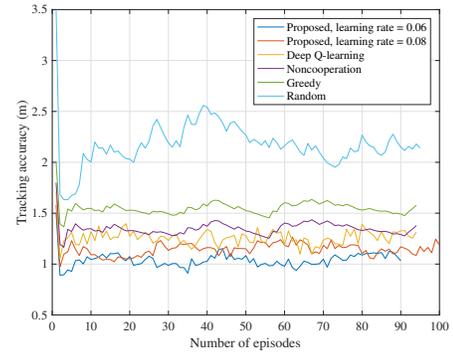}}
\caption{Tracking accuracy versus time intervals.}
\label{fig11}
\end{figure}

Fig. 10 gives the trade-off between the prediction accuracy and system cost compared with different scheduling schemes. In numerical value, the proposed LTDRA scheduling algorithm reduces the system energy consumption by the (44.0\%, 38.8\%, 21.4\%, 8.3\%), (44.4\%, 37.5\%, 30.0\% ,9.1\%), (48.7\%, 42.8\%, 33.3\%, 16.6\%), when compared with the random selection scheme, the greedy scheduling scheme, the non-cooperative scheme, and the deep Q-learning scheme considering the different prediction error upper bounds (1.5m, 2m, 2,5m). The results illustrate that the proposed scheme can reduce extra system energy consumption evidently and guaranteeing the tracking accuracy simultaneously. Our proposed intelligent scheduling scheme can significantly guarantee the real time tracking accuracy with the minimal system energy consumption. 
\begin{figure}[!t]
\captionsetup{justification=raggedright,singlelinecheck=false}
\centerline{\includegraphics[width=.75\columnwidth,]{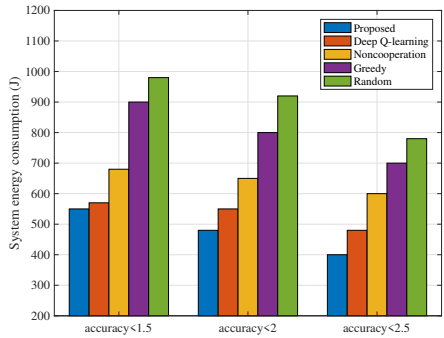}}
\caption{The tracking accuracy versus system energy consumption.}
\label{fig11}
\end{figure}

\begin{figure}[!t]
\captionsetup{justification=raggedright,singlelinecheck=false}
\centerline{\includegraphics[width=.75\columnwidth,]{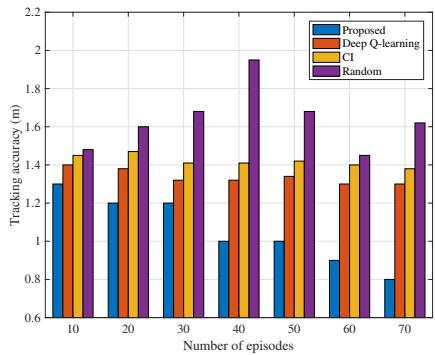}}
\caption{The tracking error versus number of iteration.}
\label{fig11}
\end{figure}   

\begin{figure}[!t]
\captionsetup{justification=raggedright,singlelinecheck=false}
\centerline{\includegraphics[width=.75\columnwidth,]{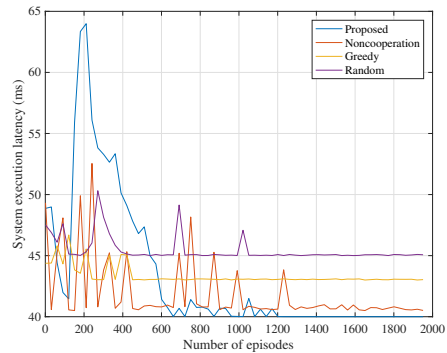}}
\caption{The system execution latency versus number of iteration.}
\label{fig11}
\end{figure}   
    
    Fig. 11 shows the tendency on system prediction accuracy with the number of the iteration. It can be observed that the system prediction error decreases as the number of iterations increases. Compared with the Centralized Implementation (CI) which adopts multi-model Bernoulli filter \cite{Alsh}, our proposed scheme can always obviously outperform during iterative process, and can reduce average 22.5\% prediction error approximately. Compared with the above five benchmarks, our LTDRA algorithm performs more stable convergence in the MTT network. In terms of unilateral indicator, our algorithm can reduce average 30\% system energy consumption to guarantee long-term target tracking. Moreover, an average 22.5\% enhancement in prediction error ensures accurate target prediction and efficient tracking performance. Considering multiple indicators, approximately 25\% energy is saved based on the same prediction error level. The LTDRA can also reduce the system response latency for exploring the optimal node scheduling strategy rapidly. On the whole, the validity of our algorithm is confirmed through multidimensional comparisons.  
    
Figure. 12 shows the performance comparison in terms of system execution latency as the number of iterations increases. Based on the proposed hierarchical target tracking structure, the system execution latency under our LTDRA scheme can be significantly reduced compared with other benchmarks. The designed intelligent cloudlet pattern provides the sufficient computing resource to respond real-time and accurate prediction for invading trajectory. Based on the numerical analysis, our proposed scheme can reduce approximately 5\%, 10\%, and 13\% response latency compared with the non-cooperative scheme, the greedy scheme, and the random scheme, respectively.

\section{Conclusion}
In this paper, we investigate the MTT-WSN system for accurate and consecutive target tracking. We design a hierarchical target tracking structure to facilitate the sensing and computing process with edge intelligence technology. The structure can achieve collaborative computing in the proposed intelligent cloudlet. Based on the design, a multi-objective optimization problem is formulated to obtain the optimal node scheduling strategy. To solve the problem, a long-term dynamic resource allocation algorithm is proposed to obtain the optimal node scheduling policy. The simulation results reveal that our algorithm can acquire the quick convergence with low response latency. Besides, our proposed algorithm can significantly enhance the tracking accuracy and decrease execution cost as well. The structure also provides a feasible approach for battery-powered MTT-WSN systems.

\bibliographystyle{IEEEtran}
\bibliography{bibs}

\end{document}